\documentclass{rQUF2e}

\usepackage{epstopdf}
\usepackage{subfigure}

\theoremstyle{plain}

\theoremstyle{definition}

\theoremstyle{remark}

\usepackage{multirow}
\begin{document}

\title{{\textit{Neural Network-based \\Automatic Factor Construction}}}


\author{Jie Fang$^{\sigma}$$\dag$\thanks{$^\sigma$First author.
Email: fangx18@mails.tsinghua.edu.cn}, Jianwu Lin$^{\ast}$$\dag$\thanks{$^\ast$Corresponding author. Email: lin.jianwu@sz.tsinghua.edu.cn}, Shutao Xia$\beta$, Zhikang Xia$\beta$, Shenglei Hu,\\ Xiang Liu$\beta$ and Yong Jiang$\dag$\\
\affil{$\dag$Tsinghua-Berkeley Shenzhen Institute, Tsinghua University, Shenzhen, China
\\$^{\beta}$Tsinghua Shenzhen International Graduate School, Shenzhen, China
\\$^{\dag}$Sino-UK Blockchain Industry Institute, Guangxi University, Guangxi, China}}

\maketitle

\begin{abstract}
Instead of conducting manual factor construction based on traditional and behavioural finance analysis, academic researchers and quantitative investment managers have leveraged Genetic Programming (GP) as an automatic feature construction tool in recent years, which builds reverse polish mathematical expressions from trading data into new factors. However, with the development of deep learning, more powerful feature extraction tools are available. This paper proposes Neural Network-based Automatic Factor Construction (NNAFC), a tailored neural network framework that can automatically construct diversified financial factors based on financial domain knowledge and a variety of neural network structures. The experiment results show that NNAFC can construct more informative and diversified factors than GP, to effectively enrich the current factor pool. For the current market, both fully connected and recurrent neural network structures are better at extracting information from financial time series than convolution neural network structures. Moreover, new factors constructed by NNAFC can always improve the return, Sharpe ratio, and the max draw-down of a multi-factor quantitative investment strategy due to their introducing more information and diversification to the existing factor pool.
\end{abstract}


\begin{classcode}C23, C33, C45, C53, C63\end{classcode}

\section{Introduction}
\subsection{Background}
In quantitative trading, predicting future returns of stocks is one of the most important and challenging tasks. Various factors can be used to predict future returns based on financial time series, such as price, volume, and the company’s accounting data. Usually, researchers define the factors that are constructed from price and volume data as technical factors, and the other factors, which are constructed from the company’s accounting data, are defined as fundamental factors. Typically, researchers conduct manual factor selection based on traditional and behavioural finance analysis. After Sharpe (1964) proposed the single-factor model for stock returns, Fama and French (1993) proposed the Fama-French Three-factor Model, which selected three important factors that could provide the most important major information to explain the stock return. Fama (2015) proposed the Fama-French Five-factor Model, which added two more factors into their Three-factor Model. Currently, some commercial multi-factor risk models, such as the Barra and Axioma model, manually select factors that are updated periodically by human researchers. However, there are two limitations in manual factor selection. First, it is very time-consuming to update factors manually. Second, it is not easy to construct certain nonlinear factors from high-dimensional data. Thus, to address data from a rapidly developing financial market, both academic researchers and quantitative investment managers have paid more and more attention to automatic financial factor construction tools. In the computer science field, this task is defined as Automatic Feature Construction, Krawiec (2002).

Feature construction in computer science is a process that discovers missing relationships between features and outcomes and augments the space of the features by inferring or creating new features. In this sense, a financial factor is a special case among general features in computer science. During this process, new features can be generated from a combination of existing features. A more straightforward description is that the algorithms use operators, hyper-parameters and existing features to construct a new feature. Sometimes, both feature construction and feature selection occur in the same procedure. Dash (1997) summarized these methods, which consist of wrapping, filtering, and embedding. Filtering utilizes only some criteria to choose a feature, and sometimes it can help us to monitor the feature construction process. It is easy to conduct but achieves poor performance. Wrapping performs well by directly applying the model’s results as an objective function. Thus, it can treat an individually trained model as a newly constructed feature. However, a considerable amount of computational resources and time is required. Embedding is a method that uses generalized features and a pruning technique to select or combine features, which serves as a middle choice between filtering and wrapping.

Statistical learning is a common factor construction tool that is used for financial factor construction, which is similar to embedding in feature construction. Harvey et al. (2016) proposed a new method for estimating latent asset pricing factors that fit the time-series and cross-section returns by using Principal Component Analysis (PCA), as proposed by Wold (1987). Feng et al. (2020) proposed model selection procedures to select among many asset pricing factors by using cross-sectional LASSO regression, which was introduced by Tibshirani (1996). However, this statistical learning mainly focuses on factor selection from existing factor pools but does not construct brand-new factors.

\subsection{Literature Review}
The most well-known and frequently employed method for automatic brand-new factor construction is Genetic Programming (GP), which is a type of wrapping method in feature construction that uses reverse polish expressions to construct new factors by an evolutionary process. However, different domains require different objective functions, and the input data structure can differ. Thus, it is very important to design this task within a specific domain. This method has been proved to work well in many industries, such as object detection, according to Lillywhite (2013), and in the educational industry, according to Romero (2004). However, its drawback is that the constructed formulas are very similar and have highly correlated outputs. The financial factor construction task uses GP to conduct the evolutionary process of formulaic factors, according to Thomas (1999) and Ravisankar (2011). WorldQuant published 101 formulaic alpha factors, which are also constructed by using this method, Kakushadze (2016). However, it only constructs a large number of similar factors, which do not contain much new information.

With the development of deep learning, more and more researchers have begun to use neural networks to extract information from raw data, and then, they add a fully connected layer to reshape the output. Some researchers have designed some specific loss functions according to specific domain knowledge. Examples are Zhang (2019) and Zhang (2020), who designed a vector-error function and derived an implicit-dynamic equation with a time-varying parameter in a neural network structure to solve the disturbed time-varying inversion problem. Yang Zhong deployed the Convolution Neural Network (CNN) to construct facial descriptors, and this method achieves better accuracy than its benchmark. Many researchers directly treat a trained neural network as a newly constructed feature. There are many representative studies in this field. For example, Shan (2017) conducted experiments on this task and deployed a deeper and wider CNN. Hidasi (2016) used a Recurrent Neural Network (RNN) to pre-locate the factor-rich region and construct more purified features. In a text classification task, Botsis (2011) leveraged the RNN to build a hierarchy classifier for text data, in which each classifier represents a part of the text. Lai (2015) proposed a network structure that uses both an RNN and a CNN to extract text information. The produced features contain more information than the previous work. With the help of a neural network’s strong fitting ability, highly informative factors can be produced by tailoring the network structure for different industries. There is no conclusion on whether using a neural network to represent a new feature can achieve better prediction performance. However, using an individual neural network to represent a new feature can add more interpretability, which is highly valued in the finance, medical, and educational sectors.

Some researchers have begun to use neural networks to give an embedding representation of a financial time series. More specifically, Feng (2019) leveraged LSTM to embed various stock time series and then used adversarial training to make a binary classification on the stock’s return. Sidra (2019) adopted a well-designed LSTM to extract factors from unstructured news data and, then, formed a continuous embedding. The experiment result shows that these unstructured data can provide much information and they are very helpful for event-driven trading. Most of these research studies focus on predicting single-stock returns by neural networks using their own time series. However, on financial factor construction tasks, we do not find literature that provides a full solution for automatic factor construction by using neutral networks.

\vspace{-0.4cm}
\subsection{Contributions}
In this paper, a novel network structure called Neural Network-based Automatic Factor Construction (NNAFC) is proposed, which can use deep neural networks to automatically construct financial factors. Different from previous research, we make three contributions in this paper for financial factor construction. (1) We create a novel loss function that differs from the accuracy of the stock’s return by using the Rank Information Coefficient (Rank IC) between stock factor values and stock returns. (2) We define a new derivable correlation formula for the Rank IC calculation to make back preparation training of neural networks possible. (3) We adopt pre-training and model pruning to add up enough diversity into the constructed factors, which helps to produce more diversified factors. NNAFC has outperformed many benchmarks in all aspects. Furthermore, different pre-training networks for prior knowledge are equipped to further improve their performance in real-world situations.
\vspace{-0.5cm}

\section{Definition of the Factor Construction Task}
\subsection{Definition of an alpha factor}
The alpha factor is a raw forecasting vector that has a certain correlation with the financial assets’ future returns. Richard (1999) pointed out that the informed expected return $E(r|g)$ can be defined as the expected return conditional on an alpha factor $g$ at time $T$. In formula (1) and (2), $r$ represents the asset return in the future, $E(r)$ represents the consensus return, $E(g)$ represents the expected forecast, $Cov(r,g)$ means the covariance between $r$ and $g$, and $Var(g)$ represents the variance of $g$.

\begin{equation}
E(r | g)=E(r)+\frac{Cov(r,g)}{Var(g)} \times (g-E(g))
\end{equation}

\begin{equation}
E(r | g)-E(r)=\frac{Cov(r,g)}{Std(r)Std(g)} \times Std(r) \times \frac{g-E(g)}{Std(g)}
\end{equation}

The excess return of the informed expected return over the consensus return is the so-called alpha of the active investment. The goal of the active investment is to maximize the alpha, which can be rewritten in formula (3).

\begin{equation}
Alpha=IC \times \sigma \times Z_{score}(g)
\end{equation}

where $IC = correlation(r, g) = \frac{Cov(r,g)}{Std(r)Std(g)}$, $Z_{score}(g) = \frac{g-E(g)}{Std(g)}$, and $ \sigma=Std(r)$. The Information Coefficient (IC) is defined as the correlation between the future asset returns and the current value of the alpha factor. Since the standard deviation of the future asset returns $\sigma$ cannot be impacted by our selection of the alpha factors and the $Z_{score}(g)$ is more of a trade-off between the excess return and the investment’s capacity, maximizing the alpha is type of equivalent to maximizing the IC.

\subsection{Definition of the factor construction process}
In cases where the alpha factor consists of many sub factors, Qian (2007) pointed out that maximizing IC is aimed at optimizing the following objective function in formula (4).

\begin{equation}
    \mathop{\max}_{v} \quad  v^{T}\widetilde{IC}-0.5\lambda v^{T} \Sigma_{\widetilde{IC}}v
\end{equation}

where $v$ is the weight vector of the sub-factors in the alpha factor,
$\widetilde{IC}$ is a value vector of the sub-factors, $\Sigma_{\widetilde{IC}}$ is the IC-value co-variance matrix of the sub-factors, and $\lambda$ is a penalty coefficient for the volatility of sub-factor IC. The optimal weight can be written as

\begin{equation}
v^{*}=\frac{\widetilde{IC}}{\lambda \Sigma_{\widetilde{IC}}}
\end{equation}

The optimal IC can be written as

\begin{equation}
IC^{*}={(\frac{\widetilde{IC}}{\lambda \Sigma_{\widetilde{IC}}})}^{T} \widetilde{IC}
\end{equation}

The optimal $IC$ shows that the goal of the factor construction problem is to construct new sub factors that have a lower correlation with existing factors (smaller elements in $\Sigma_{\widetilde{IC}}$ vectors) but a higher correlation with the financial assets’ future returns (larger elements in $\widetilde{IC}$ matrix). Thus, the factor construction problem can be defined as the following optimization problem in formula (7) below. Note that $\lambda$ in formula (4) can be ignored because it is a fixed number.

\begin{equation}
    \mathop{\max}_{FC} \quad {(\frac{\widetilde{IC_{combine}}}{\Sigma_{\widetilde{IC_{combine}}} })}^{T} \widetilde{IC_{combine}}
\end{equation}

where $\widetilde{IC_{combine}}$ is the IC vector for a set of sub-factors called $Factor_{combine}$, which consists of an existing factor set $Factor_{old}$ and a newly constructed factor set $Factor_{combine}$. Here, $\Sigma_{\widetilde{IC_{combine}}}$ is the IC co-variance matrix of this set of sub-factors.

Factor construction (FC) is defined as a process for constructing new factors from all financial data sets $FD$ and an existing factor set $Factor_{new}$ in a set of mathematical presentations {MP} to maximize the objective function in formula (7). The new factor set can be written as

\begin{equation}
Factor_{combine}=FC({FD},{Factor_{old}},{MP})
\end{equation}

Based on this definition, manual factor construction stands for a factor construction process to search potential factors in closed-form mathematical presentations by leveraging human experts’ knowledge in economics and behavioural finance research. The whole process is not completely automatic and quantitative. GP stands for a factor construction process to search for potential factors in closed-form mathematical presentations created from a set of mathematical operations based on certain genetic programming algorithms. The whole process becomes completely automatic and quantitative, and it imitates how a human constructs a mathematical formula from a set of operations. However, the set of mathematical presentations is still limited by all mathematical presentations that can be constructed by the set of mathematical operations. If the new factor must be represented by a higher order of non-linearity or fractal factions, it is difficult to accomplish by GP. NNAFC stands for a factor construction process to search for potential factors in all parameters of a neural network structure based on a network optimization algorithm to maximize IC and reduce the correlation with existing factors. The whole process becomes completely automatic and quantitative and can cover all mathematical presentations. According to Leshno (1993), the deep neural network is proved to be able to represent all types of mathematical formulas if the deep neural network has sufficient depth and width. Thus, the {MP} of NNAFC is a super set of manual factor construction and GP. In the following sections, we discuss the details of GP and NNAFC.

\section{Framework Introduction}
\subsection{Baseline method}
Before we introduce our NNAFC, we must introduce its baseline method of Genetic Programming (GP) as a benchmark. GP leverages reverse polish expression to represent factors. Each tree is an explicit formula. We use GP to conduct its evolutionary process, such as adding, deleting, merging, evolving and selecting.

In each round, the samples are selected by a pre-defined objective function. This objective function is the Spearman Coefficient between the factor value and factor return, which is exactly the same as the objective function in NNAFC. GP adds diversity into the constructed factors by changing certain parts of a reverse polish expression. In an explicit formula, a small change in the operators can make the meaning of the factor totally different. Thus, in GP, the parent and child samples can have nothing in common. As a result, the child sample might not successfully inherent some good characteristics from its parent samples, such as a good ability to predict a future stock return. Thus, genetic programming is not a good method for constructing new factors, due to its inefficient evolutionary process on this task.

GP adds diversity into the constructed factors by changing certain parts of a reverse polish expression. For example, we use a binary tree to present initial factor 1 shown in formula (9) and then conduct an evolution process via GP. In the evolutionary process, GP can make a small change in factor 1. Afterward, we can obtain a newly constructed factor 2, shown in formula (10).

\begin{equation}
Factor1=\frac{high\,price-low\,price}{volume}
\end{equation}

\vspace{-0.2cm}

\begin{equation}
Factor2=\frac{high\,price-volume}{volume}
\end{equation}

As can be seen from factor 1, it tells the relative strength of the price compared with the volume, which can be explained from the perspective of economic principles. However, factor 2 is totally different from factor 1, and it is very difficult to explain. As a result, the parent factor and child factor have little in common. The parent factor has a high IC, but the child factor might not successfully inherent the good characteristics from its parent factor. As a result, we think that GP is not a good method for constructing new factors, due to its inefficient evolutionary process on this task. The direction of its evolution and good characteristics, such as high IC, are not seen to be converging together quickly. Thus, GP on this task does not conduct an evolutionary process. It is more like a searching process, which aims at finding good solutions for an optimization problem, as shown in formula (7). GP's searching process is shown in Figure 1.

\begin{figure}[htbp]
\vspace{-0.3cm}
\centering

\includegraphics[width=13cm]{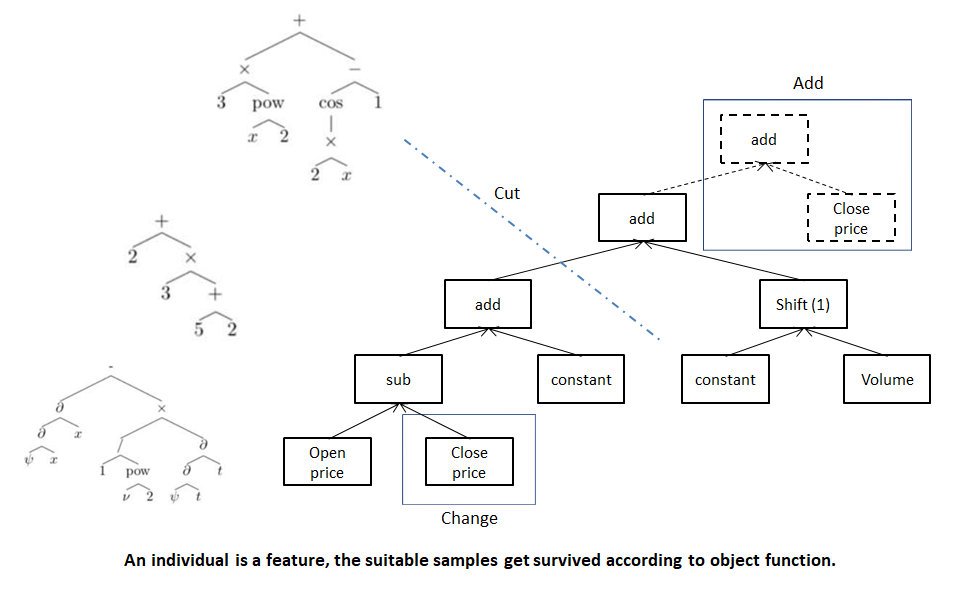}

\caption{This is the GP's evolution process. It uses reverse polish expression to represent financial factors. We can merge two formulas together to get a new formula or change a single operator in the formula to make it different.}
\vspace{-0.4cm}
\end{figure}

\subsection{Neural network-based automatic factor construction}
To overcome these drawbacks of GP, as mentioned in section 3.1, this paper proposes the Neural Network-based Automatic Factor Construction (NNAFC), a process based on a tailored neural network structure that can automatically construct financial factors toward good characteristics, such as high IC. It also diversifies the constructed factors based on prior knowledge, as represented by the variety of neural network structures. Thus, it can find good solutions efficiently for optimization problems, as shown in formula (7). The major characters of this novel network structure include the following: 1. NNAFC uses Spearman Correlation to serve as a loss function, which mimics common human practices of quantitative investment. 2. A meaningful derivable kernel function is proposed to replace the un-derivable operator $rank()$. 3. The network is pre-trained with many financial descriptors called Prior Knowledge (PK). They serve as initial seeds, which can improve the diversity of the newly constructed factors.

In NNAFC, we pre-train the neural network with an existing financial factor each time, which performs like a seed that can provide diversity to the system. According to the neural network’s universal approximation theorem, in Leshno (1993), technically, the neural network can approximate with arbitrary precision any measurable function from one finite dimensional space to another. By pre-training, the neural network can inherit good characteristics from its parent factor. With the guidance of the objective function, the new factor should have better IC than the old factor. Thus, we believe that the neural network can provide us with a more efficient and intuitive evolution process than GP on this task. The NNAFC framework is shown in Figure 2.

\begin{figure}[htbp]
\vspace{-0.3cm}
\centering

\includegraphics[width=13.5cm]{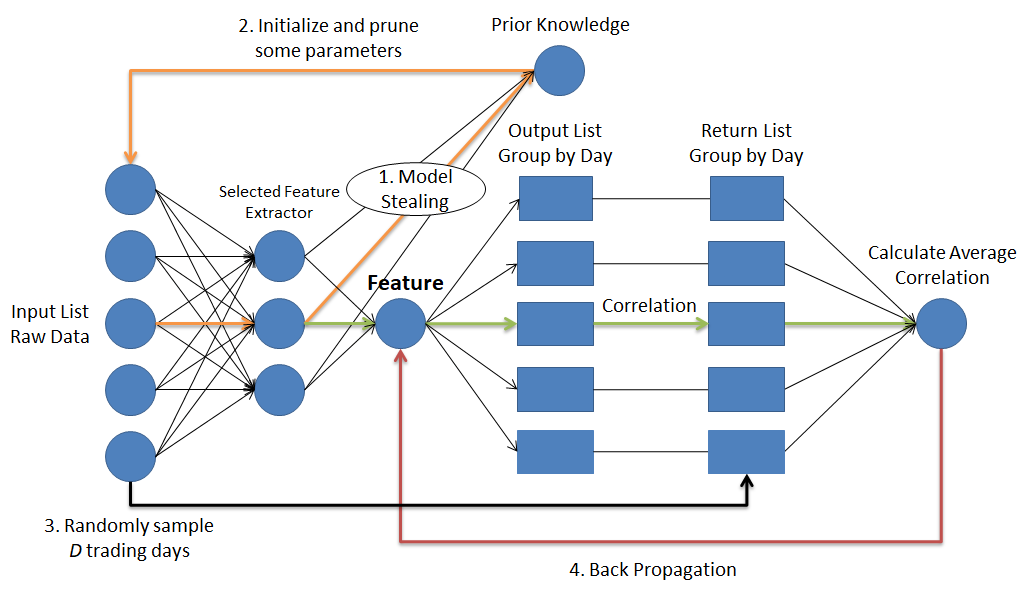}

\caption{Neural Network-based Automatic Factor Construction’s framework}
\vspace{-0.2cm}
\end{figure}

We design the batch sampling rules in which all of the stocks in the same trading day are grouped into one batch, because NNAFC focuses on predicting the relative return of the stocks on the same trading day, rather than its absolute return. More specifically, for a daily multi-factor trading strategy, we care only about the relative cross-sectional performance. Additionally, we long the relatively strong stocks and short the relatively weak stocks in each trading day to make a profit. The input tensor's shape is $(n,5,m)$, because there are $n$ stocks in each trading day, and 5 types of time series, which are the open price, high price, low price, close price and volume. The input length of each time series is $m$. In Figure 2, the output tensor of this network is called a feature list and consists of factor values with the shape $(n,1)$. Here, we assume that all of the selected networks in Figure 2 are Multi-layer Perceptrons (MLPs), for which it is easy to give a general mathematical description. In a later section, the experimental results are based on more complicated and diversified networks. $W_{i}$ is the kernel matrix in the $i\,th$ layer, $b_{i}$ is the bias matrix in the $i\,th$ layer, and $a_{i}$ is the activation function in the $i\,th$ layer. The initial layer starts from 1, and there are $p$ layers in total. The factor return tensor, with the shape $(n,1)$, consists of the returns with which we can earn from the assets if we hold the asset for a period of time. The length of the holding time is $\Delta
t$, and the current state is $t$. According to this principle, the factor value $x$ and factor return $y$ can be written as in formula (11) and formula (12).

\begin{equation}
  \vspace{-0.3cm}
  \begin{split}
  x&=l_{p}=a_{p}({W_{p}}^{T}l_{p-1}+b_{p}),\\
  l_{1}&=a_{1}({W_{1}}^{T}Input+b_{1}).
  \end{split}
  \end{equation}

\vspace{-0.5cm}
\begin{equation}
y=Factor\,Return=\frac{close \, price_{t+\Delta t}}{close \, price_{t}}-1
\end{equation}

The Spearman Correlation uses the operator $rank()$ to remove some of the anomalies in financial time series, and $rank()$ is not derivable, which is not acceptable for back-propagation in the training of neural networks. Thus, we need a derivable kernel function $g(x)$ to replace $rank()$.

\begin{equation}
g(x)=\frac{1}{1+exp(-p*\frac{x-\bar{x}}{2*std(x)})}
\end{equation}

Formula (13) standardizes $x$ into a normal distribution that is zero-centralized. Next, when the hyper-parameter $p$ equals 1.83, it ensures that 2.5\%-97.5\% of the data will lie in the range of $[mean-2std, mean+2std]$. For example, one outlier $x_i=\bar{x}+2std(x)$, and $\frac{g(x_i)-g(\bar{x})}{g(\bar{x})}\leq\frac{x_i-\bar{x}}{\bar{x}}$, and thus, the result is $std\leq0.362\bar{x}$. This finding means that if one distribution’s standard deviation is large, and it is larger than $0.362\bar{x}$, then $g(x)$ can shorten the distance between the outliers and the central point. If the distribution’s standard deviation is very small, then $g(x)$ will make it worse. However, even in this case, we can ensure that 95\% of the points are in the range $[mean-2std, mean+2std]$, which is acceptable. To show how this kernel function works, we also performed a sensitivity analysis of the hyper-parameters. Different hyper-parameter sets can represent different situations. In some stock indices, stocks’ attributes are very different. However, in others, the difference is very low. The formula focuses on cutting out the extreme part.

\begin{figure}[htbp]
\vspace{-0.3cm}
\centering

\includegraphics[width=12.5cm]{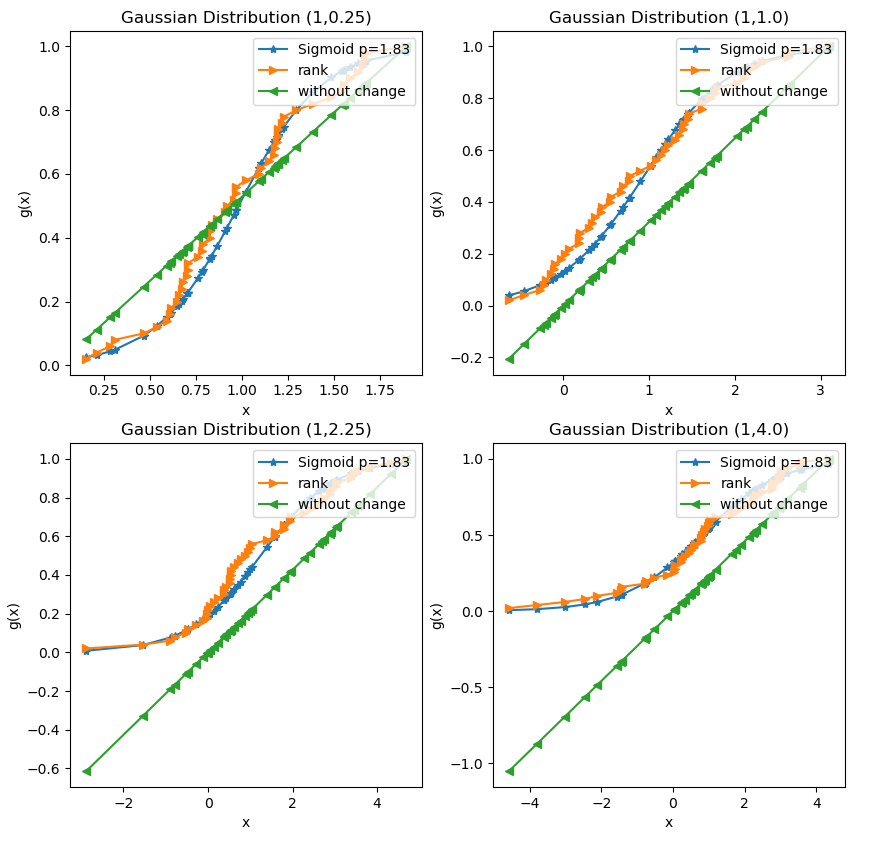}

\caption{We simulate 4 Gaussian distributions that have different mean and deviation.}
\vspace{-0.2cm}
\end{figure}

Shown in Figure 3, if the distribution’s deviation is very small, there is no need to use $g(x)$, but using $g(x)$ will not make things worse. If the distribution’s deviation is very large, then $g(x)$ can perform almost the same function as the operator $rank()$, and they can bring positive impacts.

With the kernel function $g(x)$, we define the final objective function in formula (14), where $E(x)$ represents the expected value of $x$, and $\bar{x}$ represents the average value of $x$. In each batch, we calculate the average correlation in $q$ trading days, which can make the optimization process more robust.

\begin{equation}
\vspace{-0.3cm}
\begin{split}
IC(x,y)&=\frac{E(g(x)-\bar{g(x)},g(y)-\bar{g(y)})}{E(g(x)-\bar{g(x))}E(g(y)-\bar{g(y)}))},\\
Loss&=-\frac{1}{q}\sum_{i=1}^{q}IC(x_{i},y_{i}).
\end{split}
\vspace{-0.3cm}
\end{equation}

\subsection{Putting prior knowledge into pre-training networks}
We can use pre-training to initialize a neural network. Combining pre-training and pruning on input data can improve the signal's diversity because they introduce prior knowledge into the network. Usually, the prior knowledge is represented by certain financial indicators in mathematical function formats. Model stealing is a process of fitting a mathematical function based on a given input x and output y. However, its technique is not always helpful for learning a distribution without tailoring the network structure. If we have a fixed network structure but have no idea about the target distribution, techniques such as removing the outliers (for continuous prior knowledge) and using high-temperature in knowledge distillation (for discrete prior knowledge) can improve the performance of this process. In this paper, we use some classical technical indicators, such as MA, EMA, MACD, RSI, and BOLL, and some typical financial descriptors are selected as prior knowledge for the pre-training, as shown in Table 1.

\begin{table}[htbp]
\centering
\caption{Here are the formula of some classical technical indicators and financial descriptors. They serve as prior knowledge for ADNN. Close refers to stock close price, volume refers to stock volume, and AdjClose refers to adjusted close price.}
\begin{tabular}{cl}
\hline
Technical Indicator  & Mathematical Expression\\
\hline
\multirow{1}{*}{MA}
&$MA_{N}(x_{n})=\frac{1}{N}\sum_{k=0}^N x_{n-k}$\\
\hline
\multirow{1}{*}{EMA}
&$EMA_{N}(x_{n}) = \frac{2}{N+1} \sum_{k=0}^\infty(\frac{N-1}{N+1})^{k}x_{n-k}$\\
\hline
\multirow{1}{*}{MACD}
&$MACD = EMA_{m}(i) -EMA_{n}(i)$\\
\hline
\multirow{2}{*}{PVT}
&$PVT(i) = PVT(i-1)+volume(i)*$\\
&$(close(i)-close(i-1))/close(i-1)$\\
\hline
\multirow{2}{*}{TOP10}
&$MA10 = MA\left(Close,10\right)$\\
&$TOP10 = \frac{MA10}{MA10_{top10\%}}-1$\\
\hline
\multirow{4}{*}{DC}
&$H = MA\left( High \times AdjClose/Close,n\right)$\\
&$L = MA\left( Low \times AdjClose/Close,n\right)$\\
&$M = \frac{1}{2}\left(H + L\right)$\\
&$DC = AdjClose / M$\\
\hline
\multirow{5}{*}{BOLL}
&$StdV = MStdv\left(Close,n\right)$\\
&$Mean = MA\left(Close,n\right)$\\
&$LB = Mean - Stdv$\\
&$BBL = \frac{LB}{Close}$\\
&$MStdv_{n,t} = Stdv\left(Close_{t-n:t}\right)$\\
\hline
\end{tabular}
\label{tab:plain}
\end{table}

We use $f(x)=a(w^T x+b)$ to embed the input data. The data is embedded by MLP, composed of several fully connected layers with tanh and relu activation functions. The number of neurons in each layer should be decided by the length of the input time series, where $w$ is the kernel matrix, $b$ is the bias matrix, and $a$ is the activation function. Then, we use this embedded layer to mimic the prior knowledge. In this part, we use the min(mean squared error) as the objective function, which is shown in formula (15).

\begin{equation}
    \vspace{-0.2cm}
    \mathop{\min}_{a,b,w} \ \ \frac{1}{N}\sum_{i=1}^N (y_i-f(x_i))^{2}
    \vspace{-0.2cm}
\end{equation}

In this paper, we have proposed many network structures. Different network structures require different input data structures. Some of them require time series data, but others require cross-sectional data on many stocks in the same trading day. Different input structures must be used to represent the data. However, the general rule for all of the network structures is that in each back-propagation, the input data should have all the stocks in the same trading day. To better learn the prior knowledge from technical indicators, we propose two types of network structure to learn the prior knowledge. Some technical indicators leverage only the operations on time series data, such as $SMA=\frac{1}{m}\sum_{i=1}^m close_{t-i}-\frac{1}{n}\sum_{i=1}^n close_{t-i}$. For this type of technical indicator, MLP can efficiently learn its prior knowledge. Its network structure is shown in Figure 4.

\begin{figure}[htbp]

\centering

\includegraphics[width=13cm]{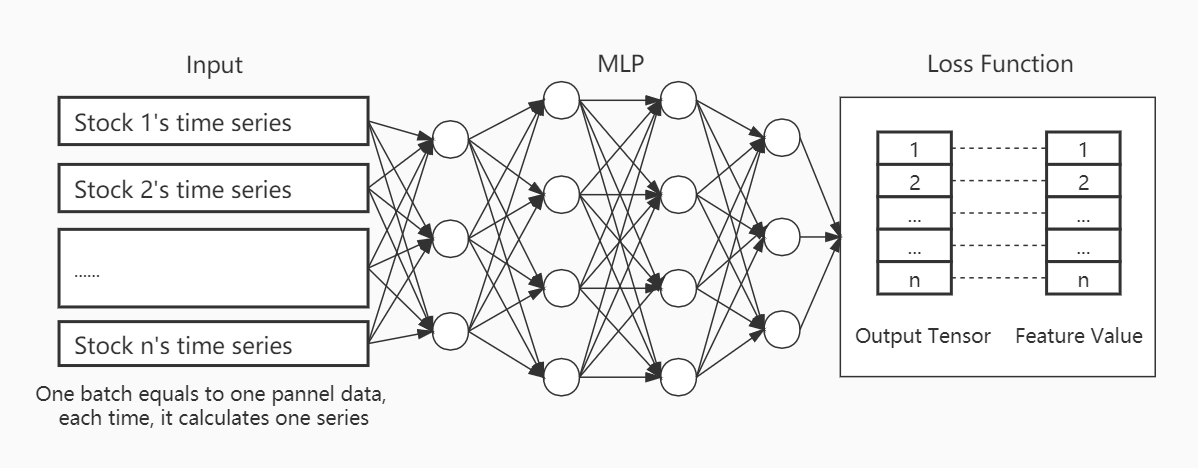}

\caption{Pre-training Structure for Time Series Factors}

\end{figure}

For other technical indicators, which require an operation on the cross-sectional data, the common MLP structure is not helpful. For example, some technical indicators need the data from both stock 1 and stock 2. Stock 1 and stock 2 are independent samples, which means that they cannot share the data with each other. Before the calculation, we do not know what parts of the data in stock 1 are needed by stock 2. To automatically and freely learn this relationship, we propose a new network structure.

\begin{figure}[htbp]

\centering

\includegraphics[width=13cm]{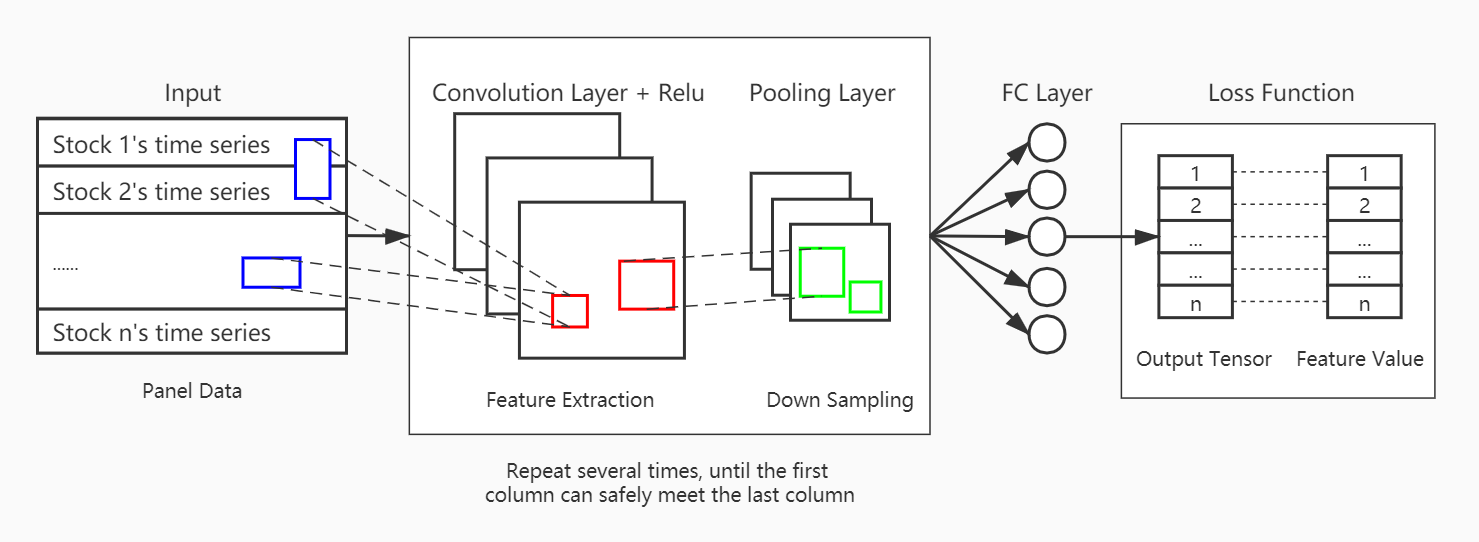}

\caption{Pre-training Structure factors with cross-sectional data}

\end{figure}

Shown in Figure 5, we leverage convolution kernels to calculate the data that belongs to different stocks. Since one kernel's size is truly small, it cannot calculate the samples at a far distance. However, if we repeat the convolution layer several times, all of the stock's data can have the opportunity to obtain the relevant data. Actually, this structure can also be used to learn time series technical indicators. In addition, this structure is not tailored for only time series. The previous structure can better mimic the performance and costs less training time. In the following experiment, these two types of network structures are both called Fully Connected Networks (FCNs).

For a deep neural network, almost all of the technical factors can be easily learned. Here, MSE or MAE cannot represent the real pre-training performance, because all of the factor values are truly small, which makes all MSE values very small. To have a better measure of the performance, $\frac{1}{N}\sum_{i=1}^N abs(\frac{y_i-f(x_i)}{y_i})$ is used to measure its error rates. The error rates of SMA, ROC, RSI, BOLL, FORCE, and other typical technical indicators are selected as prior knowledge for the pre-training. Some indicators with different parameters, such as RSI(6,12) and RSI(7,21), will be regarded as different prior knowledge because they can provide more diversity in the factor construction. We have performed experiments for a long period of time, to test whether our networks have successfully learned the prior knowledge. The training period is from Jane 2009 to March 2018, and the testing period is from May 2018 to March 2020. As mentioned above, the common MLP structure cannot address cross-sectional factors. Thus, in the following experiments, we run the pre-training process on the two structures and choose the better case. As mentioned above, although low MSE is not sufficient to prove that we have successfully mimicked the prior knowledge, it can show that the in-sample training loss and out-of-sample testing loss are small and close to each other, which is helpful in illustrating the consistent nature of the algorithm’s out-of-sample performance.

\begin{table}[htbp]
\caption{Pre-training performance, measured with the MSE of the neural network output and technical indicator}
\centering
\begin{tabular}{ccc}
\hline
Factor  &Train  &Test \\
\hline
MA    &0.00240    &0.00324\\
EMA    &0.00371    &0.00635\\
MACD    &0.00685    &0.01392\\
RSI    &0.16262    &0.16525 \\
Top10    &0.00291    &0.00286 \\
DC    &0.04314    &0.05257 \\
BOLL    &0.05336    &0.05582 \\
PVT    &0.00357    &0.00352 \\
\hline
\end{tabular}
\label{tab:plain}
\end{table}

To better prove that we have successfully learned the prior knowledge, we use these factors to make binary buy/sell decisions. If the learned trading decisions also match the real trading decisions given by prior knowledge, then we can strongly prove that it learned the prior knowledge. The accuracy of the binary classification is shown in Table 3.

\begin{table}[htbp]
\caption{Pre-training performance, measured with accuracy for binary classification.}
\centering
\begin{tabular}{ccc}
\hline
Factor  &Train  &Test \\
\hline
MA    &93.92\%    &93.48\% \\
EMA    &91.66\%    &87.81\% \\
MACD    &92.82\%    &93.77\% \\
RSI        &95.28\%    &89.11\% \\
Top10        &95.09\%    &91.25\% \\
DC        &98.05\%    &91.47\% \\
BOLL        &94.53\%    &86.75\% \\
PVT       &90.99\%    &84.75\% \\
\hline
\end{tabular}
\label{tab:plain}
\end{table}

As shown in Table 2 and Table 3, the MLP can learn prior knowledge with approximately 90\% out-of-sample accuracy. We do not require it to learn 100\% of the knowledge because 90\% of the knowledge is sufficient to serve as a source of diversity. However, why is pre-training with prior knowledge needed? Zhang (2016) pointed out that according to the concept of multi-task learning, pre-training can permanently keep some parts of the domain knowledge in the network. The different domain knowledge can increase the diversity. Pruning can filter out noise from the neural network and retain the prior knowledge that we need. In addition, the pruning rate should be controlled. A larger pruning rate will bring too much difficulty for pre-training to converge to the final optimization direction, but a smaller pruning rate could lose the diversity. Frankle and Carbin (2018) pointed out that the ideal pruning rate should be approximately 0.2-0.5, and the mask matrix is composed of only 0 and 1. All of the settings are the same as those in the paper of Ding (2015). After embedding the data as $f(x)$ in formula (16), we obtain its parameter matrix $W$. Then, we create a mask matrix to prune the parameters. For example, $x_{ij}$ in the parameter matrix is relatively small, which means that some of the input data is useless. Then, $M_{ij}$=0 is set to permanently mask this value. If the $x_{ij}$ is not useless, then we set $M_{ij}$=1. This method can retain the diversity in the network. Furthermore, it can focus on improving the current situation, without heading into an unknown local minimum. The pruning process by using matrix $M$ is shown in formula (16):

\begin{equation}
    f(x)=(W\cdot M)^\mathrm{ T }x+b
\end{equation}

After pre-training and pruning the network, we use the objective function shown in formula (14) for the NNAFC’s training. We simply reshape the input data into a graph, and then, we use the Saliency Map proposed by Simonyan (2014) to look at how the raw data contributes to the final constructed factor. The Saliency Map's calculation method is shown in formula (17). If we assume that the input pixel is I, then the Saliency Value can be approximated by the First Order Taylor Expansion, $S(I_{p})\approx w^{T}I_{p}+b$. After calculating the derivative on the variable $I$, we can obtain the contribution rate of pixel $I_{p}$.

\begin{equation}
    W_{I_{p}}=\frac{\partial S(I_{p})}{\partial I_{p}}
\end{equation}

\begin{figure}[htbp]

\centering
\vspace{-0.2cm}
\includegraphics[width=15cm]{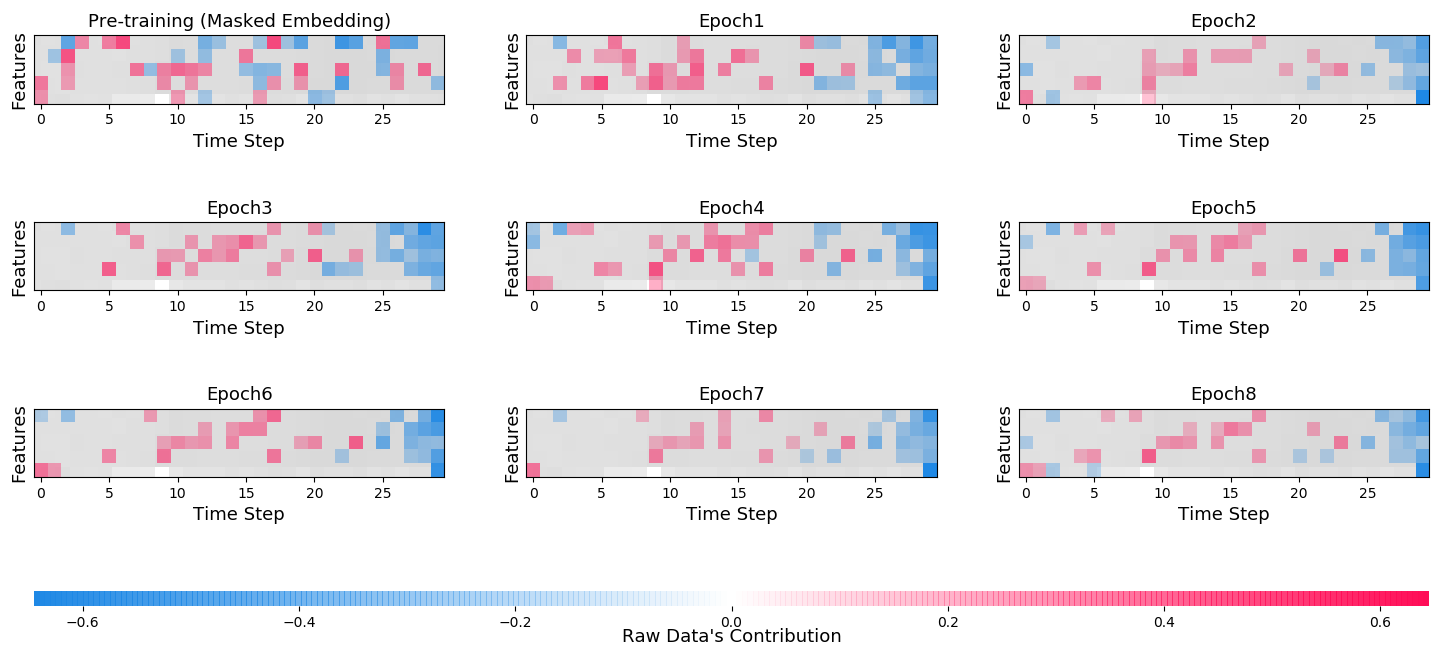}

\caption{Factor construction process of one stock series for illustrating the evolution process of NNAFC. However, in our factor construction process, we do use cross-sectional data as inputs.}
\vspace{-0.2cm}
\end{figure}

The training process is illustrated in Figure 6. The y-axis is [open price, high price, low price, close price, volume]. The x-axis is the time step of an input time series. The red colour means a positive contribution, and the blue means a negative contribution. The colour's depth has a positive correlation with the contribution rate. From Figure 6, we can obtain further understanding of the construction process. First, there is a reverse technical factor, according to the raw data's negative contributions at the latest time. Second, we can know that this factor is mainly constructed by the price information, but not volume information. Third, the components of this factor are discrete. The factors constructed by human experts are usually continuous. This difference makes the factors constructed by NNAFC uncorrelated to manually constructed factors by human experts. Because the trading opportunities are limited, if we all use similar trading signals, the trading will be more and more crowded. Prior knowledge can be viewed as seeds for new factors, which can bring diversified and robust initialization. Starting from this prior knowledge would be a very safe choice. Furthermore, as mentioned in section 3.1, one of GP's shortcomings is that the child factor might not inherit the good characteristics of its parent factor. However, for NNAFC, all of the updates in the parameters are dominated by back-propagation. The mechanism of back-propagation is gradient descent, which ensures that every optimization step is continuous. These continuous characteristics can ensure that the parent factor and child factor have a lot in common. Thus, GP is more similar to an advanced searching algorithm, but NNAFC is more like an advanced evolutionary algorithm.

We also conduct experiments on different feature extractors (in the computer science field, different feature extractors mean different neural network structures) to have a better understanding of the pre-training on different financial time series. There are two motivations for conducting experiments on different feature extractors. First, different feature extractors require different input data structures. We explored many different methods for organizing the input data. We conducted experiments on many basic networks and SOTA (state-of-the-art) networks, including Fully connected network (FCN), Le-net, Resnet-50, Long Short Term Memory Network (LSTM), Transformer (because Transformer is too large, we used only the self-attention encoder) and Temporal Convolution Network (TCN). FCN has 3-5 hidden layers, with 64-128 neutrals in each layer. We used the tanh and relu function to serve as the activation function, and the output layer should not be activated. For Le-net, Resnet-50, Transformer, and TCN, we did not change their structures, but we employed only a smaller stride because the input data has a low dimension compared with the real picture.

The second motivation is that different feature extractors have their own advantages and disadvantages. Some of them aim at extracting temporal information, but the others aim at spatial information. Some of them are designed for a long term series, but others are designed for quick training. We believe that they can make our factor pool even more diversified.

\subsection{Summarizing the pros and cons}
Until now, we have introduced all of the network settings of NNAFC, and we have shown the characteristics of NNAFC compared with GP. Here, we summarize their pros and cons. First, GP conducts a very low efficiency evolution process, and its performance is more similar to a searching process. As a result, the factors constructed from GP are very similar. However, NNAFC can conduct an efficient evolutionary process, and it fully inherits the diversity from the prior knowledge. Thus, the factors constructed by NNAFC are more diversified.

Second, the factors constructed by GP are formulaic factors, which is easier to explain. However, for NNAFC, although we attempt to explain it in section 3.4, its explainable ability is still limited.

Third, for some simple factors, GP is truly easier to explain. However, the simple factors cannot provide much information. In other words, there is no room for us to improve both the quality and quantity together when the factors are very simple. Only when the formula is very complex, we have the possibility to reproduce many different and useful factors based on it. However, if the formula is very complex, we cannot understand and explain the factors constructed by GP, either. In this situation, both methods cannot be easily explained. The non-formulaic factors constructed by NNAFC could be better. At least, this approach has the ability to retain more information. As mentioned in section 3.1, a deep neural network has the ability to represent any formulas.
\newpage

\section{Numerical Experiments}
\subsection{Experiment setting}
In the experiment part, we performed two types of experiment. First, we conducted experiments to compare the performance of NNAFC and GP from the perspective of IC and diversity. Second, we conducted experiments to measure their contribution in real trading strategies. We compared a strategy that uses only the prior knowledge, a strategy that uses both prior knowledge and the factors constructed by GP, and a strategy that uses both prior knowledge and the factors constructed by NNAFC. This comparison can show how much contribution our method can bring to the current multi-factor strategy.

We used daily trading data in the Chinese A-share stock market (in the following, we call it the A-share market data), including the daily open price, high price, low price, close price and trading volume over the past 30 trading days. The raw data is standardized by using its time-series mean and standard deviation. Both the mean and standard deviation are calculated from the training set. We attempted to use these inputs to predict the stock return in the next 5 trading days, which stands for the weekly frequency and is a common period used in short-term technical indicators. In the A-share market, we cannot short single stocks easily because of the current limitations in stock lending, but we can short some important index futures.

We performed some research on selecting some reasonable hyper-parameters. During the model training process, we calculated the average value of IC over 20 randomly selected trading days. For each experiment, 250 trading days served as the training set (no technical factor can work well for a long period of time), and the following 30 trading days served as the validation set, while the following 90 trading days served as the testing set. These hyper-parameters are common settings in industrial practices. To make a fair comparison, the same setting is deployed for the GP algorithm.

In this paper, we analyse the construed factors’ performance from different perspectives. Based on the definition of alpha factors, we use the information coefficient (IC), shown in formula (6), to measure how much information is carried by a factor. For diversity, each factor value should be normalized at first. The softmax function in formula (18) can eliminate the effect from the scale without losing the factors’ rank information.

\begin{equation}
Softmax(x_{i})=\frac{\exp(x_{i})}{\sum \exp(x_{i})}
\end{equation}

Then, the cross-entropy is used to measure the distance between two different factors’ distributions on the same trading day.

\begin{equation}
Distance(f_1,f_2)=\sum softmax(f_1) log \frac{1}{softmax(f_2)}
\end{equation}

In formula (19), $f_1$ and $f_2$ refer to different factors’ distributions in the same trading day. K-means is used to cluster the distance matrix of the relative distance between two factors. The average distance between each cluster centre refers to the diversity of factors on this trading day. We have shown in formula (6) that a higher optimal IC for an investment strategy is related to a higher sub IC and a larger factor diversity. In addition to measuring the IC and diversity, the performance of a trading strategy based on the constructed factors is also measured, such as absolute return, max-drawdown, and sharp-ratio.

To put all in a nutshell, we summarize the construction procedure mentioned in section 3. And then, we illustrate the entire process of constructing a new factor via NNAFC. First, we pre-train a neural network with a technical indicator MA as prior knowledge. The pre-training performance has been well illustrated in Tables 2 and 3, and we can pre-train this factor with 93.92\% accuracy. Second, we train the neural network by maximizing IC defined in formula (14). During the back-propagation, the neural network has been changed, and the output factor value changes also. This process has been shown in Figure 6. Because we use the IC to serve as an objective function and the mechanism of back-propagation is gradient descent, the newly constructed factors will have higher IC than the initialized factor. We also use the following experiment results to prove this assumption. At last, the trained neural network is a newly constructed factor. As mentioned in the section 3.4, it has pros and cons compared with the benchmark. We definitely cannot know its closed-form formula; however, we can know what raw data contributes to it and how different it is compared with the initialized factor, shown in Figure 6.

\subsection{Beating the state-of-the-art technique}
NNAFC can be equipped with different neural networks. In this test case, NNAFC is equipped with 4 layers in a fully connected neural network (FCN). The experiment shows that NNAFC can beat the GP. We propose three schemes to help illustrate NNAFC’s contribution and to show how it beat the GP. The \emph{only GP} means only using GP, \emph{Only NNAFC} means only using NNAFC to construct factors, and \emph{GP and NNAFC} means using the GP’s value to initialize NNAFC and then construct factors. The out-of-sample results of the experiments are summarized in Table 4.

\begin{table}[htbp]
\caption{The performance of different schemes}
\centering
\begin{tabular}{ccc}
\hline
Object  & Information Coefficient & Diversity \\
\hline
\emph{Only GP}       & 0.094  & 17.21     \\
\emph{GP and NNAFC}       &0.122  & 25.44      \\
\emph{Only NNAFC}    & 0.107  & 21.65     \\
\hline
\end{tabular}
\label{tab:plain}
\end{table}

\emph{Only NNAFC} is better than \emph{Only GP}, which means that NNAFC outperforms GP on this task. We also find that \emph{GP and NNAFC} is the best, which means that our method can even improve on the performance of GP. However, in real practice, we should leverage the constructed factors to form a multi-factor strategy and compare its performance with GP. The specific strategy setting is the same as in section 3.4, and we have repeated this experiment for different periods of time. The long-term back-testing result is shown in Table 5.

\begin{table}[htbp]
\centering
\caption{Strategy’s absolute return for each scheme.}
\begin{tabular}{ccccc}
\hline
Time                                                            & \emph{Only GP} & \emph{GP and NNAFC} & \emph{Only NNAFC} & \emph{ ZZ500}    \\
\hline
\begin{tabular}[c]{@{}c@{}}Train:2015.01-2015.12\\ Test: 2016.02-2016.03\end{tabular} & +2.59\%  & +5.74\%  & +4.52\%  & +1.67\%  \\
\hline
\begin{tabular}[c]{@{}c@{}}Train:2016.01-2016.12\\ Test: 2017.02-2017.03\end{tabular} & +5.40\%  & +10.26\% & +8.33\%  & +2.53\%  \\
\hline
\begin{tabular}[c]{@{}c@{}}Train:2017.01-2017.12\\ Test: 2018.02-2018.03\end{tabular} & -5.27\%  & -4.95\%  & -4.16\%  & -6.98\%  \\
\hline
\begin{tabular}[c]{@{}c@{}}Train:2018.01-2018.12\\ Test: 2019.02-2019.03\end{tabular} & +13.00\% & +15.62\% & +15.41\% & +13.75\% \\
\hline
\end{tabular}
\label{tab:my-table}
\end{table}

As shown in Table 5, the \emph{Only NNAFC} always has better performance than the \emph{Only GP} during the long-term backtest. The results show that our method has also beaten the SOTA in real practice. However, will there be more powerful feature extractors to discover knowledge from financial time series? And what shall be the suitable input data structure for different financial time series? We attempt to answer these questions in the next few sections.

\subsection{Comparing different feature extractors}
For the hardware equipment, we used 20 g GPU (NVIDIA 1080Ti) and 786 g CPU (Intel Xeon E5-2680 v2, 10 cores). Based on this setting, we show the amount of time that we need to construct 50 factors. Moreover, the time to restore 50 trained networks and obtain their factor values will be substantially faster than traditional factors. Because some of the traditional factors are constructed with complicated explicit formulas, these formulas are not suitable for matrix computing. Using neural networks to represent factors in matrix computing, which have faster testing speeds. For the overhead of this framework, the time complexity is acceptable. With 20G GPU, it costs only several hours to construct 50 factors. However, the GPU memory is a large problem. In each back-propagation, we need to store more than 3000 stocks' time series, each stock has at least 5 time series, and each time series contains more than 100 points. Thus, this framework requires at least 20G GPU in resources for the Chinese stock market's backtest, and more GPU resources will be better.

\begin{table}[htbp]
\centering
\caption{The higher the information coefficient (IC) and diversity are, the better is their performance. Normally, a good factor’s long-term IC should be higher than 0.05.}
\begin{tabular}{ccccc}
\hline
Type & Network & IC & Diversity & Time\\
\hline
Baseline &GP& 0.072 &17.532 &0.215 hours    \\
\hline
Vanilla &FCN &0.124 &22.151 &0.785 hours           \\
\hline
\multirow{3}{*}{Spatial}
& \multicolumn{1}{c}{Le-net} &0.123 &20.194 &1.365 hours \\
&Resnet-50 &0.108 &21.403 &3.450 hours \\
\hline
\multirow{3}{*}{Temporal}
&LSTM &0.170 &24.469 &1.300 hours \\
&TCN &0.105 &21.139 &2.725 hours           \\
&Transformer &0.111 &25.257 &4.151 hours    \\
\hline
\end{tabular}
\label{tab:plain}
\end{table}

Shown in Table 6, $Type$ means the category of neural networks. For example, both Le-net and Resnet are designed for extracting spatial information. Allen (1999) pointed out that all neural networks can produce more diversified factors than using GP. For Le-net and Resnet, they do not provide us with more informative factors, but for more diversified factors, there is LeCun (1999) and He (2016). Temporal extractors are especially better at producing diversified factors, such as LSTM and Transformer, Hochreiter (1997) and Vaswani (2017). For TCN, Dea (2018) proves its ability to capture the temporal rules buried in data. However, they have enormous differences. TCN relies on the CNN, but LSTM and Transformer still contain an RNN. Normally, the transformer uses an RNN to embed the input data. The existence of a recurrent neural network structure can contribute to more diversity. All of the neural networks mentioned above can produce more informative and diversified factors than GP. For the same types of networks, the results suggest that the simple network structure performs relatively better than the sophisticated networks.

\subsection{Real-world use case test}
In the real-world use case test, we use the factor constructed via NNAFC. At the same time, we also use the factors constructed by human experts. This approach should give a fair setting for a comparison, and we want to see whether the NNAFC can bring marginal benefits for the traditional and mature investment system. Table 7 shows the back-testing results of our strategy.

In the training set, the stocks whose returns rank in the top 30\% in each trading day are labelled as 1, and the stocks whose return ranked in the last 30\% of each trading day are labelled as 0. We abandon the remaining stocks in the training set, according to Fama (1993). After training these factors with Chen (2015)'s XGBoost using binary logistics mode, we can obtain a trained model that can make binary predictions. The prediction result reflects the odds as to whether the stock’s return will be larger than 0 in the following several trading days. It defines the 50 factors constructed by human experts as \emph{PK 50}, and the 50 factors constructed by NNAFC as \emph{New 50}. In another case, in the training set, we use XGBoost to train 100 factors, which are composed of \emph{PK 50} and \emph{New 50}. Then, we select 50 factors whose factor weights are relatively large among these 100 factors. The weights can be calculated by the frequency of using the factor. For example, in a tree-based algorithm, the important features can obtain more opportunity to serve as a splitting feature. By using this approach, we can use XGBoost to select 50 important factors, which consist of both traditional factors and neural network-based factors. We want to stress the point again that the feature selection process occurred in the training set. We do not conduct feature selection in the validation set or testing set. Furthermore, during each backtesting, this factor selection process should only be conducted once. Last, we define the selected 50 factors as \emph{Combine 50}, which represents the practical use of our factors.

\begin{table}[htbp]
\centering
\caption{The investment target is all Chinese A-share stocks, except for the stocks that cannot be traded during this period of time. The strategy’s commission fee is 0.3\%.}
\begin{tabular}{cccccc}
\hline
Type                                          & Target                       & Group       &Revenue &MD &SR\\ \hline
\multicolumn{1}{l}{\multirow{5}{*}{Baseline}} & ZZ500                        & Stock Index & 19.60\% & 13,50\%       & 1.982        \\ \cline{2-6}
\multicolumn{1}{l}{}                          & HS300                        & Stock Index & 18.60\% & 20.30\%      & 1.606        \\ \cline{2-6}
\multicolumn{1}{l}{}                          & PK                           & PK 50       & 24.70\% & 18.90\%      & 2.314        \\ \cline{2-6}
\multicolumn{1}{l}{}                          & \multirow{2}{*}{GP}          & GP 50       & 17.60\% & 25.30\%      & 1.435        \\ \cline{3-6}
\multicolumn{1}{l}{}                          &                              & Combine 50    & 25.40\% & 14.80\%      & 2.672        \\ \hline
\multirow{1}{*}{Vanilla}                      & \multirow{1}{*}{FCN}                                    & Combine 50 & 29.60\% & 15.70\%      & 3.167        \\ \hline
\multirow{2}{*}{Spatial}                      & \multirow{1}{*}{Le-net}                                 & Combine 50 & 27.50\% & 16.40\%      & 2.921        \\ \cline{2-6}
                                              & \multirow{1}{*}{Resnet-50}                 & Combine 50 & 29.30\% & 17.20\%      & 2.787        \\ \hline
\multirow{3}{*}{Temporal}                     & \multirow{1}{*}{LSTM}                                   & Combine 50 & 29.90\% & 15.00\%      & 3.289        \\ \cline{2-6}
                                              & \multirow{1}{*}{TCN}                      & Combine 50 & 26.90\% & 16.80\%      & 2.729        \\ \cline{2-6}
                                              & \multirow{1}{*}{Transformer}
                                                                  & Combine 50 & 27.20\% & 15.10\%      & 2.806        \\ \hline
\end{tabular}%
\end{table}

As shown in Table 7, \emph{HS300} and \emph{ZZ500} are important stock indices in the A-share stock market. The strategy for the excess return is the annualized excess return of the long portfolio vs. the index. The max drawdown is the worst loss of the excess return from its peak. The Sharpe ratio is the annually adjusted excess return divided by a certain level of risk. These indicators can show the strategy’s performance from the perspective of both return and risk.

In a multi-factor strategy, a higher correlation among the factors will reduce the strategy’s performance, and it is shown in the definition of the factor construction process in formula (7). The goal of factor construction is not to find factors with higher performance, but to find factors that can improve the overall performance of the combined factors selected from \emph{PK 50} and \emph{New 50}. Thus, combining the factors from both the new and existing human experts’ factors is more reasonable and suitable for practical use cases. In all cases, our \emph{combined 50} is better than \emph{PK 50} and GP's \emph{Combine 50}, which means that the NNAFC can construct more useful factors than GP with a reasonable factor selection process.

\subsection{Comprehending the results}
In section 4, the numerical experiment results show that the LSTM can extract more information than FCN. We suspect that only the RNN and FCN are helpful for extracting information from the financial time series. To verify this idea, we constructed 50 factors by using an FCN, 50 factors by using a spatial neural network and 50 factors by using a temporal neural network. Then, we clustered all of these factors into three groups by using k-means. The goal of this process is that there are mainly three types of neural networks, and we want to find whether the constructed factors have a similarity relationship. The definition of the distance has been mentioned in formula (19) in section 4.1. To visualize this distance matrix, this matrix should be transformed into a 2D graph. We initialize one of the cluster centres as $(0,0)$ and then determine the other two cluster centres according to their relative distances and a given direction. This direction will influence only the outlook of this graph and will not influence the shared space between the two different clusters. For samples that belong to the same cluster, their location is determined according to the relative distance between a cluster centre and a randomly generated direction. As a result, we can obtain the distribution of the constructed factors. The larger the factor's sparsity is, the larger the contribution of a certain type of network. Sometimes, the factor's sparsity is low, but if it has never been overlapped by the other network's factors, its distinct contributions are also highly valued. The experiment results are shown in Figure 7.

\begin{figure}[htbp]
\vspace{-0.2cm}
\centering

\includegraphics[width=13cm]{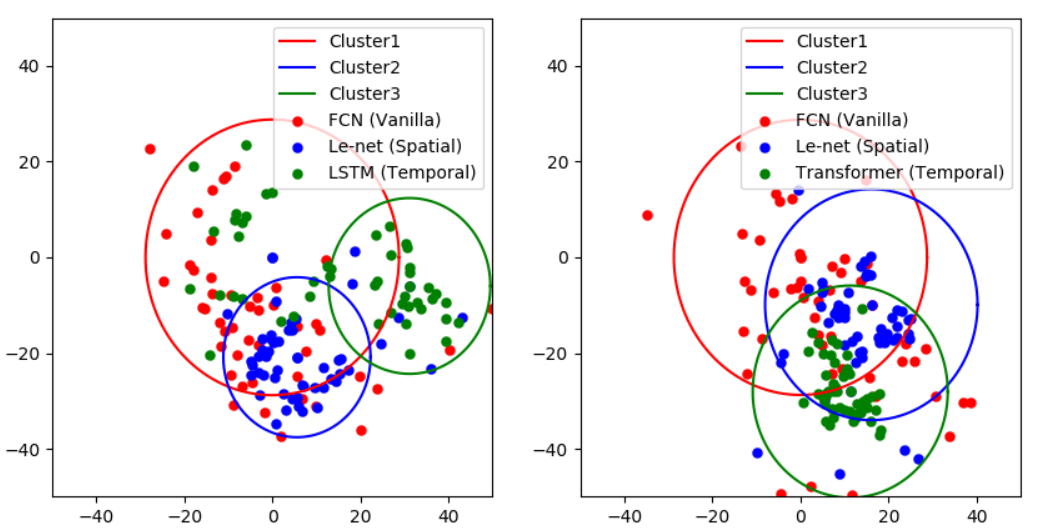}

\caption{Cluster different networks (spatial against temporal)}
\vspace{-0.2cm}
\end{figure}

As shown in Figure 7(left), the factors constructed by the LSTM have the sparsest distribution, which means that the network structure that focuses on temporal information is excellent at extracting diversified information from the financial time series. However, a large space is shared by FCN and Le-net. We can regard Le-net’s information as a subset of the FCN. Combined with the CNN’s poor performance in sections 4.2 and 4.3, it looks such as that the CNN structure does not make a substantial contribution in extracting information from the financial time series. Figure 7(right) is an extra experiment, whose results support this conclusion as well.

Except for the type of neural network, actually, the complexity of the neural networks also influences the result. We provide extra experiments in Figure 8. Normally, the model’s complexity should meet the dataset’s complexity. Most of the financial decisions are still made by humans, and thus, the signals are mostly linear and simple because this way is how the human brain processes information. A very complicated network structure will bring in extra risk of over-fitting.

\begin{figure}[htbp]
\vspace{-0.2cm}
\centering

\includegraphics[width=13cm]{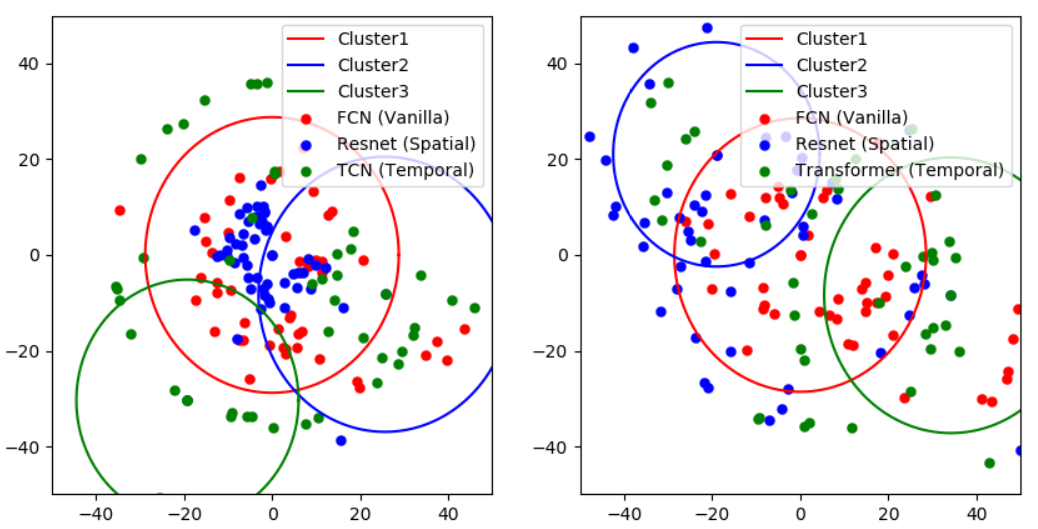}

\caption{More complicated neural networks compared with the networks used in figure 8.}
\vspace{-0.2cm}
\end{figure}

In Figure 8, the shared space between Vanilla and Spatial is still larger than the space shared between Vanilla and Temporal. Thus, it appears that the reason why the LSTM's information coefficient can outperform the other networks is that the recurrent structure and fully connected structure are truly helpful in extracting information from financial time series. At the same time, these two structures focus on a different part of the information, which makes their joint effort more valuable. Using an RNN to obtain an embedding representation of a financial time series is the best choice, as shown in our experiments. However, a more complex time-decay structure could be used, such the TCN, but the Transformer does not perform better than the LSTM. We believe that the trading signals in the Chinese A-share market are mostly linear, and thus, a very complex non-linear feature extractor (a neural network structure) is not suitable at present.

However, while the stock market is developing, more and more investors crowd into this game. We think that the factor crowding phenomenon will become more and more clear. In addition, as more and more tradings are made by algorithms, the non-linear part in the trading signals will be larger. Thus, for quantitative trading, we believe that the complicated and tailored neural network structure will have its supreme moment in the near future.

\section{Conclusions and Future Research}
In this paper, we propose Neural Network-based Automatic Factor Construction (NNAFC). This framework can automatically construct diversified and highly informative technical indicators with the help of prior knowledge and different feature extractors. In both numerical experiment and real-world use case tests, it can perform better than the state-of-the-art in this task, which is genetic programming. Although different network structures perform differently in this factor construction task, they can contribute to the diversity of the factor pool. Thus, they are also highly valuable for the multi-factor quantitative investment strategy. Furthermore, we also conduct experiments to comprehend their contributions and differences. For further research, this framework can also be tested on a company's fundamental and market news data.

%
%
%
\bibliographystyle{splncs04}
%

\end{document}